\documentclass{article}

\usepackage{graphicx}
\usepackage{amssymb}
\usepackage{amsmath}

\usepackage{epstopdf}

\usepackage{textcomp}

\usepackage{hyperref}

\usepackage{amsbsy}

\usepackage{float}

\usepackage{wrapfig}

\input{epsf}

\usepackage{cancel}

\usepackage{mathpazo}

\usepackage{geometry}

\usepackage{slashed}
\usepackage{xcolor}

\begin{document}
\def\xy{(x, \, y)}
\def\xyz{(x, \, y, \, z)}
\def\cH{\cal H}
\def\H{Hamiltonian }
\def\EL{\cal{EL}}
\def\eq{\end{quotation}}
\def\bq{\begin{quotation}}

\def\bqs{{\small \begin{quotation}}
\def\eqs{\end{quotation}}}
\raggedbottom

\def\pex{\paragraph{Exercise:}}

\def\bn#1\ea{\begin{align} & #1\end{align}}

\def\bq{\quotation}

\def\ab{\allowbreak}
\def\nn{\nonumber}
\def\ra{\rightarrow}
\def\d{\mathrm{d}}
\def\bf{\textbf}
\def\diag{\mathrm{diag}}
\def\dim{\mathrm{dim}}
\def\D{\mathrm{D}}
\def\E{\mathrm{E}}
\def\integer{\mathrm{integer}}
\def\tr{\mathrm{tr}}
\def\cA{\cal A}
\def\cD{\mathrm{\cal D}}
\def\cF{\cal F}
\def\cL{\cal L}
\def\cO{\cal O}
\def\s{\,\,\,}
\def\a{\alpha}
\def\b{\beta}
\def\e{\epsilon}
\def\f{\phi}
\def\fr{\frac}
\def\g{\gamma}
\def\h{\hat}
\def\i{\index}
\def\p{\partial}
\def\x{\xi}
\def\AA{\vec A}
\def\BB{\vec B}
\def\EE{\vec E}
\def\NN{\vec \nabla}
\def\JJ{\vec j}
\def\DD{\Delta}
\def\xx{\vec x}
\def\eps{\epsilon_{0}}

\def\ba#1\ea{\begin{align}#1\end{align}}

\def\bit{\begin{itemize}}
\def\eit{\end{itemize}}
\def\im{\item}

\def\ov{\over}

\newcommand{\nnn}{\nn \\ &}

\newcommand{\sch}{Schr\"{o}dinger }

\def\N{Newtonian }
\def\cN{{\cal EN}}
\def\bra#1{\left< #1\right|}
\def\ket#1{\left| #1\right>}
\def\bracket#1#2{\left<#1\mid #2\right>}
\def\EV#1#2#3{\bra{#1}#2\ket{#3}}

\title{ Is Ball Lightning \\ a Signal of Magnetic Monopoles?} 

\author{John P. Ralston \\ 
Department of Physics \& Astronomy \\
University of Kansas, Lawrence KS 66045 USA} 

\maketitle

\abstract{While ball lighting is known to exist from thousands of observations, its properties have never been explained by known physics. The combined order of magnitude of power, size, time scale, and characteristic behavior of ball lightning have defeated every model. The failure of standard physics does not hinge on fine details, but on the breadth of qualitative features of the phenomenon itself. We consider the possibility that ball lighting may be a signal of physics beyond the Standard Model. The mass and energy scale of ball lightning is remarkably consistent with new physics scales of grand unified theories. We analyze a suggestion that monopole catalysis of baryon decays in air might explain ball lightning. Nothing can be done to make it consistent, including saturating interaction cross sections with unitarity bounds. However a high mass dyon partially decaying by neutralization of deeply bound electromagnetic energy explains several puzzles naturally. The study suggests new ways to search for dyons and monopoles. The organizational skills of ambitious experimental physics groups could lead to productive searches beyond the Standard Model that do not need a new particle accelerator.  }

\section{Physics Not Explained by the Standard Model}

{\it Ball lightning} is a well-documented phenomenon that has long defied theoretical explanation. It is not St. Elmo's fire, a corona discharge. It is not a chemical phenomenon of burning material, nor a plasma bubble of trapped microwaves, while such ideas have been proposed. A generally spherical object from 10-100 cm diameter is seen moving at a few meters per second in air, or sometimes bouncing off the ground or other objects. The object emits light  with a few watts of power in the visible spectrum. The phenomenon is short-lived, usually being visible for 1-10 seconds, or at most 100 seconds. It is associated with lightning, hence the name, yet not strictly correlated with lightning being nearby. Ball lightning has been repeatedly observed passing through material barriers such as doors, walls and glass: That is very significant.

Ball lighting has been observed for centuries. It is reported several hundred times per year in the US. The organizations of the military, meteorological, and government establishments take ball lightning seriously. Several books\cite{Uman1969-qj,Barry1980,Stenhoff2010-by,Boerner2019-eh} and a {\it Physics Reports}\cite{Smirnov1993-qu} article review the subject. Appendix C of Ref.\cite{Uman1969-qj} has a summary for atmospheric scientists that can be read online. Uman writes that ``Visual sightings are often accompanied by sound, odor, and permanent material damage, and hence it would appear difficult to deny the reality of the phenomenon''. Chapter 20 of a well-known ``lightning handbook'', Rakov and Uman\cite{RakovUman2003} reports the existence of almost 5000 observations of ball lightning. Stenhoff's book\cite{Stenhoff2010-by} is quite complete and also rather critical, listing many causes of misidentification, while the phenomenon cannot be made to go away. (The bibliography lists about 2400 references.) Ref.\cite{hgss-12-43-2021} presents 41 selected case study observations by scientists and trained observers. 

Yet the {\it phenomenology} of ball lightning has no conventional explanation. Attempts to make a theory explain selected features, but are inconsistent with other features. Trapped microwaves\cite{Wu2016-ju} cannot be sustained for billions of cycles without dissipating.\cite{Shmatov2019-gt}. In a {\it Physical Review} article\cite{PhysRev.135.A390}, Finkelstein and Rubenstein presented order of magnitude estimates (similar to ours), writing that ``It is concluded that the plasmoid model is impossible and that energy must be supplied to the ball during its existence if the order of magnitude of the reported energies and times are accepted''. If an independent power source would exist, several ``plasma ball'' models\cite{Shmatov2003-in,Lowke2012,Morrow2017-aw} can agree with qualitative features. Morrow's model\cite{Morrow2017-aw} is mathematically well developed. The problem is the source of power. Morrow writes that  ``Some additional mechanism must power the light output of ball lightning.'' More than a dozen models have been proposed\cite{Smirnov1993-qu,RakovUman2003}, but none explain the source of energy. 

The failure of conventional physics to explain ball lightning does not hinge on fine details. The order of magnitude of power, size, time scale, and qualitative character of the phenomenon have defeated every model. Models invoking exotic chemistry\cite{Abrahamson2002}, plasma or electromagnetic physics are ruled out by ball lighting passing unchanged through walls and glass windows. The possible signal of new physics from ball lightning now is similar to the signal of new physics seen in radioactivity around 1900. In both cases, the unexplained source of energy is decisive.

\subsection{The Monopole Hypothesis}

A recent paper by Karl Stephan\cite{stephan2024balllightningmagneticmonopoles} reviews the failure of conventional models for ball lightning, and suggests it may be a signal of exotic physics. Stephan cites a 1990 paper by Korshunov\cite{Korshunov:1990gf} that connected ball lighting to the Callan-Rubakov mechanism\cite{Rubakov:1981rg,Rubakov:1982fp,Callan:1982ac,Callan:1982au} for catalysis of baryon decay by magnetic monopoles. 

\subsubsection{Background} 

For readers not already familiar, magnetic monopoles (henceforth, ``monopoles'') have been a mainstream subject since Dirac found a condition\cite{Dirac:1931kp} involving monopole magnetic charge $g$. The relation is $eg=n'\hbar c/2$, where $e$ is the unit of electric charge, and $n'$ is an integer. The topic expanded exponentially when t' Hooft \cite{tHooft:1974kcl} and Polyakov\cite{Polyakov:1974ek} found monopoles as classical soliton solutions in gauge theories. 

The theories describe a wide range of length scales, from the macroscopically long ranged magnetic field, to internal substructure characterized by known physics and new physics. Electrically charged monopoles, called {\it dyons}, are essential to classifying monopoles. From a theoretical view uncharged monopoles may be an exception rather than a rule. From the outside in, the possible structures of a monopole include non-relativistic atomic-type bound states with known particles and nuclei. The size of such states is proportional to the inverse mass of the bound particle. States bound by their magnetic moment exist, but can ``fall to the center'' unless some form factor effect stabilizes them. Relativistic bound states of Dirac fermions are much different. Kazama and Yang\cite{Kazama:1976fm} (KY) found states with a binding energy equal to their mass. The next length scale is set by the strong interactions, where confinement of color occurs on distances of order 1 fm =$10^{-13}$ cm. The confinement scale sets the scale of the proton's mass $m_{p}c^{2}\sim 1$ GeV in natural units $\hbar=c=1$.\footnote{To convert from $\hbar=c=1$ units to SI units, multiply or divide as needed by $\hbar c\sim 2\times 10^{-5}$ eV cm, and use the numerical value of $c$.} Moving inward toward the monopole center, the electroweak interaction has a distance scale about 1000 times smaller and determined by Higgs physics. 

Higher energy physics appears at smaller distance scales. Grand unified theories (GUTs) have processes at very short distances where quarks could transform to leptons, potentially causing proton decay. Observational limits on the proton's lifetime exceed $10^{31}-10^{33}$ years, depending on mode\cite{ParticleDataGroup:2024cfk}. An energy scale of GUTs exceeding $10^{16}$ GeV is estimated from the limits on the proton lifetime. With few exceptions the 't Hooft-Polyakov type monopoles require new gauge groups or representations also required by unified theories. This explains why the core size of a magnetic monopole is believed to be very small, of order $10^{-30}$ cm, and that monopole masses are believed to be  $10^{16}-10^{18}$ GeV. 

Monopole physics was revolutionized again with predictions that GUT monopoles could catalyze baryon decay\cite{Rubakov:1981rg,Rubakov:1982fp,Callan:1982ac,Callan:1982au}. In brief, a quark making its way to the center of a monopole would be transformed to other particles. The decaying baryon would release much of its rest mass as energy, with the monopole presumably left unchanged. 

While monopole physics depends on the model, there is little debate that baryon catalysis by a GUT monopole {\it has a mechanism} to release a great deal of energy. Ball lightning is a phenomenon needing a source of energy. Hence the discussion. For more information on monopoles, Rajantie's semi-popular reviews\cite{10.1063/PT.3.3328,RajanteMonopoles} are very readable. Milton\cite{Milton:2006cp} presents a thorough review of theory. Giacomelli {\it et al} \cite{giacomelliMagneticMonopoleBibliography2000b,Giacomelli:1984gq} present extensive bibliographies. 

\subsection{Outline of the Paper}

\label{sec:almostintro}

Section \ref{sec:nogo} finds intriguing numerical coincidences between the mass scales of GUTs and the size, mass, and power scales of ball lightning. This is very interesting. Yet the calculations of Section \ref{sec:nogo} practically rules out catalysis of baryon decay in air to explain what is known. The same theory predicting baryon catalysis predicts reaction cross sections and rates in air that are much less than ball lightning needs. Regardless of those predictions, we then use the power requirements to predict a cross section. The result would violate very general unitarity bounds of quantum mechanics. (Actually the discrepancy is  ``only'' a few orders of magnitude.) 

The coincidence with GUT scale numbers remains interesting. The lifetime of ball lightning is extremely long for an object with so much energy. This suggests the decay of some high mass soliton might be the source of ball lightning's energy. Solitons are not point-like particles but extended objects. The lifetime of a metastable soliton is liable to be determined by quantum mechanical tunneling. The reason some radioactive nuclei have very long lifetimes is that tunneling can inhibit processes that would be fast in perturbation theory. 
 
Solitons in quantum field theory are subject to quantum fluctuations that tend to destabilize them. If a soliton has a long lifetime, the default is topological stability. Topological charge is a differential invariant that cannot change under smooth variations of the classical fields. It is no accident that monopoles are topological solitons in almost all models. The magnetic charge is a topological charge. By a short chain of logic, a long-lived source of GUT-scale energy indicates a high mass soliton with topological metastability, and we return to monopoles. 

Since the complete destruction of a magnetic monopole is almost taboo, we are led to explore the possibility that a monopole undergoes a transition of its internal state. We call this a ``decay'' in the same sense that the decay of an atomic level or a radioactive nucleus does not imply destruction of the object. There are several ways a monopole could decay. The puzzle of ball lightning is why lightning-related phenomena would initiate the process. Directly related is the puzzle of why an unstable monopole would survive to be observed. It turns out that the existing literature on monopoles has the information to deduce a natural mechanism, which is presented in Section \ref{sec:physicalpicture}. 

Section \ref{sec:checks} presents consistency checks, and Section \ref{sec:discussion} concludes with a discussion. We invite readers to educate themselves about ball lightning and convince themselves it is real. It would be desirable to have reproducible laboratory experiments with multiple events. Section \ref{sec:discussion} list some new suggestions in that direction. Given the information that has been reported, our premises are that ball lightning exists, and demands to be explained.

\section{Bottom-Up Analysis} 

\label{sec:nogo}

We will not be concerned with the physics of ``plasma ball'' models. Lab experiments with unlimited energy show that fireballs can be made. To reiterate, the critical requirement that models need is a source of energy to match the observations. 

We will proceed {\it from} the phenomena {\it to} a mechanism of whatever new physics is needed. Much of monopole physics is very sophisticated, and assumes specified symmetry groups, representations, symmetry breaking patterns, and exact solutions. We cannot rely a top-down approach because there are too many possibilities to provide guidance

From the bottom up, producing visible light with power of order 1W must involve $10^{2}-10^{4}$W of total power or more, and not less. For reference the chemical soliton called a ``candle flame'' produces about 100W of energy and 1W of visible light. Ball lightning has not been associated with much perceived heat. However there is evidence for ionizing and ultraviolet radiation, as directly measured and deduced by induced fluorescence\cite{Stephan2016-el}, which we discuss in Section \ref{sec:ionizing}. Since there is little heat, the spectrum of energy must not be thermal, mostly likely with a peak above the visible region. 

Can this energy scale be produced by baryon catalysis in air? Let $dE/dt$ be the rate of energy production with yield $\Delta E$ per event, cross section $\sigma$, for a particle moving with relative velocity $v$ in a medium with number density $n$, thus \ba {dE \over dt} =nv\sigma. \nn \ea To conserve angular momentum a baryon's decay products must include on odd number of less massive fermions, of which there exists only the electron $(m_{e} \sim 0.5$ MeV), muon $(m_{\mu} \sim 106$ MeV), and neutrinos. Electrons with GeV-scale energies have a range in air of about 10 meters, which is too large for the size of ball lightning. The light quark hadrons less massive than the proton are the pions ($m_{\pi}\sim 140$ MeV), a few other mesons $f_{0}, \, \rho,  \, \omega$...(masses 500 MeV, 770 MeV, 800 MeV, respectively). To a very good approximation, the other mesons decay very quickly to pions. $\pi^{0}$ decays to 2 very penetrating gamma rays with energies about 67 MeV, while $\pi^{\pm}$ decay to neutrinos and muons with enormous range. Already this does not appear promising to reproduce the phenomenon of ball lighting. 

Korshunov assumed that a monopole catalyzing $^{14}N \ra ^{13}C$ would absorb a final state $\pi^{0}$ with 100\% efficiency, breaking into 13 nuclear fragments each with 3.5 MeV of energy. While this is not typical of nuclear physics, we can entertain the positive possibility that nuclei would be disrupted into a number of nucleons with few MeV energy scales. The range in air of a 3.5 MeV proton is about 15 cm. That is in the regime of BL sizes, as Korshunov noticed, and that also summarizes his contribution. The rest of Korshunov's calculations unfortunately have several errors that invalidate its claims.\footnote{In Eq. 2 Korshunov cites a reaction rate in air of $6 \times 10^{14}s^{-1}$. The corresponding collision cross section is $\sigma_{K}\sim 1.2 \times 10^{-10} cm^{2}$. That suffices to invalidate the paper. Yet one must be wary of possible typos and check all the steps. Instead of using a cross section, Korshunov computes the rate $dN/dt =4\sqrt{2}\pi R^{2}n v$, using $R= 2.9 \times 10^{-7} \, cm$. Then $\pi R^{2} \sim 10^{-3} \sigma_{K}$ is inconsistent with the rate cited. The value of $R$ itself is very close to $n^{-1/3}$, and claimed to come from a magnetic Debye screening radius, which (if relevant) computes to $R \sim 10^{-11}$ cm. While Korshunov cites a catalysis cross section of $10^{-28}cm^{2}/\beta^{2}$ from Rubakov, no calculation uses it. A concept error of 100\% catalysis efficiency appears at 2.1, `` In the series of pioneer scientific papers, it was shown that within the framework of 'grand unification' theories, there is no place for the mechanism not favoring the monopole-nucleon interaction.'' The scale of 3.5 MeV/nucleon seems to come from knowing the size of ball lightning and computing the energy. Multiplying by 13 nucleons gives $\Delta E=45.4 MeV$ per event, which Korshunov claims comes from $m_{\pi}-97.1 MeV=138.5-97.1=45.4$ (which is true). The pion's kinetic energy is forgotten, and we cannot explain 97.1 MeV. No calculation is given for the statements ``The escaped energy of the described process is about 90 kJ/sec. The maximum value one may get is about 1.3 mJ/sec,'' which are written in a way to seem obvious.}

Accepting an overestimate, let us continue with locally deposited energy $\Delta E \sim GeV$ per decay. For air at standard temperature and pressure the nucleon number density $n \sim 8 \times 10^{20}cm^{3} \ra 10^{21} \, cm^{-3}$ and the thermal velocity $v\sim 5 \times 10^{4}\, cm/s$. With $\sigma \sim 10^{-28} cm^{2}$, to order of magnitude $dE / dt \lesssim 5 \times 10^{4+21-28} GeV /s$. Converting 1 eV $\sim 10^{-19}$ J, the estimate is 
\bn {dE \ov dt} \lesssim  5 \times 10^{-13} {\text{J} \over s}({\Delta E \ov GeV  })({n \over 10^{21}/cm^{3}} )( {\sigma \over 10^{-28}cm^{2}})  . \nn \ea Solve for the necessary cross section as a function of power $P$ in kW: \ba \sigma \gtrsim 5\times 10^{-12}cm^{2} ({ 10^{21}\, cm^{3} \ov n  })  ({  5 \times 10^{4} cm/s\ov v}) ({GeV \ov \Delta E})({P \ov 10^{3}W}) . \label{power} \ea  Delivering even 1 W of power to make BL needs $\sigma >> 10^{-15} \,  cm^{2}$ with the other factors of order 1. That is 13 orders of magnitude larger than the optimistic GUT catalysis estimates of order $10^{-28} cm^{2}$. (Fryberger\cite{Fryberger:1994vv} independently made a connection between ball lightning and baryon catalysis, which is subject to the same conclusion. Lipkin's proposal \cite{Lipkin:1983gp} for monopole induced nuclear fission and beta decay fares worse for producing even less energy.)

This might not be fatal. A theoretical bias that everything about monopoles is ``well understood'' should be avoided. Suppose the ``agent'' of reaction is some other soliton, or a variant of dark matter than has escaped detection, or a very low mass $t$-channel exchanged particle, a non-standard black hole, or something else. The unitarity bound on partial wave cross sections\cite{Sakurai2020-ji} with angular momentum $\ell$ is \ba \sigma_{\ell}={4\pi (2\ell+1)\sin^{2}\delta_{\ell}/\ov k^{2} }\lesssim {10 \over (m_{b}c^{2})^{2} (v/c)^{2}}, \ea for beam mass $m_{b}$ and most $\ell$. The wave number $k =\hbar k$. Numerically $(1/m_{b}c^{2})^{2}=4 \times 10^{-26}cm^{2}/(m_{b}c^{2}/GeV)^{2}$. Combining with Eq. \ref{power} gives \ba   10^{-12}({ 10^{20}\, cm^{3} \ov n  } )({GeV \ov \Delta E})({P \ov 10^{3}W}) \,  cm^{2} < \sigma < 10^{-15}({10^{4}cm/s \ov v})^{2}\, cm^{2} . \label{inequal} \ea It is just barely possible to get 1 W of total power this way if there is some mechanism for GeV-scale energy production at the unitarity bound.

Single fermion $s$-wave catalysis cross sections equalling the unitarity limits have actually been proposed. It is not theoretically inconsistent for the entire $J = 0$ partial wave to violate baryon number conservation\cite{Sen:1983yq}. As the Particle Data Group review\cite{Workman:2022ynf} puts it, ``catalysis is model-dependent and is not even a universal property of all GUT monopoles.'' Yet an isolated fermion's cross section is not relevant here. Quarks are confined on the strong interaction scale leading to maximal catalysis cross sections of the same scale $\sigma \sim 10^{-26}-10^{-28} \, cm^{2}$. 

The observed cases where unitarity bounds are saturated are resonant processes.\footnote{For a period ``Sommerfeld enhancements'' of arbitrary size were used to estimate hypothetical dark matter cross sections violating unitarity bounds. While an enhancement from an accumulation point in the spectrum of bound states exists, its effects are order 1 for weak coupling, and {\it consistent} with unitary bounds. See Ref\cite{Backovic:2009rw} for limits and Ref\cite{miltonCoulombResummationMonopole2008c}for an application to monopoles.} Low energy neutron-nucleus scattering has resonant cross section peaks exceeding $10^{-19} \, cm^{2}$. However the resonant peaks are very narrow, leading to large fluctuations in collision rates, not very large rates. Atomic ionization cross sections are resonant, and of order an atom's size-squared, namely $10^{-16} \, cm^{2}$. But we don't know how to get $\Delta E \sim GeV$ from an atomic process. 

These order of magnitude estimates have been pushed to a narrow zone of marginal consistency. Perhaps this is interesting. An advocate might argue that narrow limits on parameters constitutes a prediction. But a robust approach cannot depend on fine tuning. Ball lightning is a family of macroscopic phenomena somehow associated with a wide range of energy and time scales. A viable mechanism should fit all the observations, not just a narrow selection. Despite calculations that suggest giving up, there is good reason to be persistent. {\it Ball lightning is a signal of something new that actually exists.}

\subsection{The Agent Must Be Heavy} 

An episode of ball lightning ends when it either goes out of sight, fades to nothing quietly, or explodes like a firecracker or gunshot, (rarely ``a cannon''.) For reference, 1 gram of TNT delivers $4 \times 10^{3}$J of energy, and smokeless gunpower is similar. In rare cases walls have been knocked down, etc., perhaps increasing the energy scale to MJ. There is definitely a clue in a {\it range} of energies that is not a perfectly {\it sharp} energy. 

We turn to decay (i.e. a transition) of a heavy agent as a source of power. The ball lightning observations requires the object delivers of order $10^{4}$W of energy during a 1-100 s period of observation $\delta t$. Converted the power $P$ to particle physics units, \ba P\gtrsim  ({10^{19} eV \ov J})({10^{-9} GeV \ov eV}) \gtrsim 10^{14} {GeV \ov s}. \nn \ea If 1\% of the agent's mass is converted to energy in the transition, the mass is of order $m_{MM}\sim P\delta t \gtrsim 10^{16}-10^{17}$ GeV. This is a very interesting coincidence with the mass scales of many GUT theories $m_{GUT} \sim 10^{16}$ GeV. We accept this as a clue. The Planck scale of $10^{19}$ GeV might be relevant, just as well.

Suppose the agent is a monopole. The Dirac quantization condition sets its SI unit magnetic charge $g=3 \times 10^{-9}\, Nt/T$, or some integer multiple of this. The absence of observing large accelerations in the earth's magnetic field has information. Suppose the change in velocity $\Delta v \sim a \Delta t \lesssim 1$  m/s over a time period $\Delta t \sim 10 $ seconds. Then $a \lesssim 10^{-2} \, m/s^{2}$. The magnetic force on the monopole is $F_{B}=gB_{E}$, where the magnetic field near the earth's surface $B_{E} \sim 1 \, gauss =10^{-4}$ T. This puts a bound on the objects \N mass $m_{MM}$: \ba m_{MM} ={F_{B} \over a} \gtrsim 3 \times 10^{-9} \times 10^{-4}   \times 10^{1}( { m/s^{2} \ov a} {B_{E} \ov gauss} \sim 10^{-11} \, kg. \nn \ea In mass units 1 GeV $\sim 10^{-27}$ kg, and \ba m_{MM} \gtrsim 10^{16}\, GeV. \ea An object need not decay entirely. If 1\% of the mass is lost in a transition, multiply by 100, and $m_{MM} \gtrsim 10^{18}\, GeV$. This is one more coincidence with scales assumed for new physics.\footnote{We recognize that some model estimates of monopole masses are order $4\pi m_{GUT}/\a$, namely larger than a GUT scale $m_{GUT}$. We are avoiding model-dependent details.}

Suppose the agent has electric charge. The acceleration $a_{E}$ of a charge $Qe$ in an electric field $E$ is $ 10^{-19+27}(GeV/m)(Q/e)E/(V/m).$ A steady state electric field of about 100 V/m exists in the Earth's atmosphere. It is not easy to observe, because air is such a poor conductor of electricity that most objects are good conductors in comparison, and define their own equipotential surfaces. To keep $a_{E} \lesssim 0.1 m/s^{2}$ with $Q=1$ needs $M \gtrsim 10^{11}$ GeV. 

Can a monopole go through glass and wood, or must it stop? The energy losses of electromagnetic objects are strong functions of velocity. Atomic ionization losses predicted for $v \lesssim 10^{-4}c$ are very small\cite{cecchiniEnergyLossesMagnetic2016}. Elastic scattering and phonon scattering dominate. The energy loss per distance $dE/dx \sim 10^{2}- 10^{3}$ MeV/cm in liquid hydrogen\cite\cite{cecchiniEnergyLossesMagnetic2016} for $10^{-5} \leq v/c \leq 1$. The number density of targets in glass is similar to liquid hydrogen to order of magnitude. Objects of mass $10^{17}$ GeV with speeds 1-10 m/s have approximately 1-100 GeV of \N kinetic energy. Comparing these numbers, slow GUT scale monopoles could generally pass through glass, but not TeV scale monopoles. A

As for an associated plasmoid passing through an insulator, Ref. \cite{Ohtsuki1991-pg} observed microwave-generated ``plasma fireballs'' inside waveguides passing through ceramic plates 3mm thick without causing damage. Anthony {\it et al}\cite{anthonyExperimentalStudiesLight2009} present an extensive study of light emission phenomena inside superconducting RF cavities, with many photographs.

\subsection{Electromagnetic Bound States with Monopoles}
\label{sec:bound}

Information about electromagnetic bound states of monopoles will be critical to resolving the puzzles cited in Section \ref{sec:almostintro}.

Electric charges do not form non-relativistic quantum bound states with uncharged monopoles. Yet monopoles do form bound states with magnetic moments above a critical magnitude. Milton gives a comprehensive review\cite{Milton:2006cp}. For fermions with magnetic moment Hamiltonian $H=-e(1+\kappa) (e/2m)\sigma \cdot \vec B$, binding needs the anomalous coefficient $\kappa \gtrsim f(J)$, where $f(J)$ is a function of $J$ and the magnetic charge. For estimates $f(J)$ is of order 1 (unless $J$ is very large). This has been known from the 1950's\cite{PhysRev.83.899} and 1970\cite{sivers1970possible}.

While bound states are certain, binding energies are uncertain. The interaction is singular at short distance $r \ra 0$ and needs regulation. Table 1 of Ref.\cite{Milton:2006cp} gives many examples. In one estimate\cite{Bracci1984-zy} the binding energy of a proton is 15 keV. Many nuclei are more deeply bound. Form factors introduce a natural scale for a short distance cutoff, yet using the information gauge invariantly is not straightforward. Another analysis\cite{Olaussen:1984xb} estimated proton binding might be anywhere in the range 50 keV-1 GeV due to form factor effects. 

The relativistic case is curious. Solving the Dirac equation, Kazama and Yang\cite{Kazama:1976sr} found an electrically charged fermion-monopole system is unstable by ``fall to the center.'' Formally speaking the Hamiltonian is not automatically self adjoint, but needs a self-adjoint extension of boundary conditions.\cite{Jackiw:1975fn,Kazama:1976fm} Yet an anomalous magnetic moment (even for infinitesimal $\kappa$) creates stability. Stable wave functions go like $e^{-\kappa/mr}/r$ as the radius $r \ra 0$. Here the mass of the fermion $m$ sets the distance scale. Besides the non-relativistic bound states, KY find an infinite family of deeply bound states with total energy 0. That is, the binding energy equals to rest mass. KY disavowed zero energy bound states of protons due to form factor effects, without giving a calculation. The distance scale cutting off the wave function is of the same order as the smearing of magnetic moment by a form factor. KY concentrate on zero energy bound states of electrons. They suggest monopoles would be surrounded by a plasma of $e^{+}e^{-}$ pairs  produced out of the vacuum. \footnote{ It seems that $\mu^{+}\mu^{-}$ pairs, not mentioned by KY, would also be made with zero energy. At the time $\tau^{+}\tau^{-}$ had not been established. }

The need for a self-adjoint extension is not separate but intimately related to the sensitivity of baryon-monopole scattering to non-perturbative ``clouds'' of fermion-antifermion
pairs\cite{Callan:1982ah,Balachandran:1983fh,Weinberg:1998aw}. The literature has a repeated theme that a monopole might be electrically charged (a dyon) or trap an electric charge at more than one possible distance scale. One scale is the non-relativistic regime of radius $r\sim 1/m$ set by the bound particle's mass. Another scale is that of strong interactions, and another has charge in a core of size $1/m_{MM}$.  
\section{Catalysis of Monopole Decay}

\label{sec:scenario}

\subsection{A Physical Picture Guided by Data} 

\label{sec:physicalpicture} 

The observations and monopole literature suggest a physical picture: 

 \bit \im Monopoles on Earth are bound by their magnetic moments to nuclei. The same mechanism binds protons, but less deeply. Nuclei will eject bound protons and replace them. As Milton\cite{Milton:2006cp} puts it, ``Monopoles bound with kilovolt or more energies will stay around forever.''   
 
 \im An event (``lightning'') liberates the monopole from its nuclear binding, and releases it into the air. We say it is ``ionized,'' for discussion. The new monopole can form metastable states with protons. Free protons are available in air ionized by lightning. 
 
 \im Protons absorbed by a monopole can tunnel to its center at a non-negligible rate. 
 
 \im As the proton is absorbed, the energy scale of ball lightning indicates a mechanism must exist to release much more energy than catalyzing proton decay. 
  
 \im Recall that monopoles might be dyons. The electrostatic energy from charge in the deep core of a dyon should be a significant fraction of its mass. In fact, the size $r_{MM}$ of a GUT monopole is proportional to its inverse mass, $r_{MM} \sim 1/m_{MM}$, ignoring model-dependent factors. The electrostatic energy $E_{es} \sim \a /r_{MM}$, where $\a \sim 1/137$ is the fine structure constant. On the scale of the Standard Model, $\a m_{MM} \sim 10^{14}-10^{16}$ GeV is immense. In fact, it is the energy scale of ball lightning.
 
 \im Rather than catalyzing baryon decay, it will be energetically favorable for the monopole to ``eat and keep'' a proton's charge, releasing energy as the monopole decays. This will take some time, because tunneling is coupled to the conversion of monopole field energy to Standard Model quanta.
 
 \im Several puzzles are now resolved by a natural physical mechanism. One puzzle is why monopole transitions should be generic, rather than a topological accident. Electrostatic energy is generic. 
Another puzzle is why monopoles would survive in the early universe until now. Long-lived bound nuclear states shield a monopole from having its decay catalyzed. Since nuclei are composite objects and heavier than protons, their tunneling to the core would be relatively suppressed. The puzzle of why lightning would initiate a transition of a high mass object is explained. 

\im There will be many ways to ionize a monopole. Just as knocking a neutron or alpha particle out of a stable nucleus can lead to beta decay, ionization by a cosmic ray could initiate a sequence of internal state transitions. This explains how ball lighting can sometimes spontaneously appear independently of lightning. 
\im A range of dyon electric and magnetic charges naturally leads to a range of ball lightning species, output powers and lifetimes, even for a single object. We can also consider multiply magnetic-charged monopoles becoming destabilized and fragmenting into smaller units of magnetic charge. When of topological origin, multiple magnetic charges occupy distinct topological classes, which rather generally would be less stable than spherically symmetric, minimal charge classes.\cite{Gervalle2022-kk}. In fact a history of observations have seen ball lightning sometimes split into smaller ball lightnings. The quantization of electric and magnetic charge plus its conservation can be considered to predict splitting and multiple forms of ball lightning.  \eit 
 
\subsection{Theoretical Difficulties}

While much of the phenomenology of ball lightning now has a possible explanation, the time scale does not seem to be something theory can reliably estimate. Here are ideas we have considered: \bit \im Monopoles disassociating by false vacuum transition were discussed in Ref. \cite{Steinhardt:1981mm}. Ref. \cite {Callan:1983nx} uses a short-distance interaction to extinguish the topological charge of a skyrmion. Mechanisms of core instabilities of 'tHooft-Polyakov-type monopoles have long been known\cite{Bais:2002ae,striet_more_2003}. Gervalle and Volkov\cite{Gervalle2022-kk} present a thoughtful discussion of monopole stability, along with interpreting the Cho-Maison monopole\cite{Cho:1996qd} found in the Standard Model. We considered these processes, which would need some ad-hoc reason for association with lightning,  before realizing that GUT scale dyons would prefer to consume protons than catalyze them. \im Releasing energy of order $\a m_{MM}$ to standard model quanta cannot occur instantly. A static (time independent) classical field configuration is nothing like a heavy particle vibrating on the pole of its propagator. For an estimate based on dimensional grounds, consider a transition rate producing $e^{+}e^{-}$ pairs \bn \Gamma \sim \a^{2} m_{e} \sim 10^{-4} ({MeV\over 200 \, MeV\, fm})({ 10^{23} fm\ov}s) \sim 10^{17} \, s^{-1}; \nnn {dE \ov dt} \sim  \sum 10^{17} \, MeVs^{-1} =10^{4} \, Js^{-1}. \nn \ea By no means is this reliable. However it is the sort of estimate the phenomenon suggests. The physical picture suggested by ball lightning is something like a slow first order phase transition. ``Critical slowing down'' is common in phase transitions, and known to occur with first order type.  Static processes nominally have no currents and cannot radiate without some tunneling process. Tunneling through metastable configurations is indicated. \im The uncertainties of non-perturbative, semi-classical strongly interacting physics are huge. Nevertheless some approaches suggest themselves. Recall the Euler-Heisenberg\cite{Heisenberg:1936nmg} effective action and the Schwinger mechanism\cite{Schwinger:1951nm}. Dunne \cite{doi:10.1142/9789812775344_0014} reviews the history and current development. Static, translationally invariant electric and magnetic fields ($E=|\EE|$, $B=|\BB|$) fields pair produce fermions of mass $m$ at a rate \ba {dN \over d^{3}x dt} \sim {e^{2}EB \over 8 \pi^{2}}\sum_{n} \, {1\ov n}\coth({n \pi B\ov E})exp(-{ m^{2}\pi n \over eE}). \label{screen} \ea This is the case for $\EE$ parallel to $\BB $. Notice the exponential dependence on $m^{2}\pi n /eE$. An unstable region has $E \gtrsim m_{e}^{2}/e$, just as dimensional analysis suggests. Larger $E$ fields drive the exponent to 1, while the pre-factor determines the rate. We can imagine a formula generalized to a dyon might have the same qualitative trends. Vacuum polarization of dyons will exist. The effect of vacuum polarization in QED redistributes zero net charge to partly screen a central charge. Yet the rate of the process is not directly related to the rate of dyon decay by absorbing a proton charge. We stopped pursuing this because there are many ways to make what is unknown even more complicated. \im Returning to the analogy with radioactivity, there are reasons to believe that a universal theory to predict monopole decay processes is not feasible. It is much more ambitious than predicting the lifetime of an arbitrary radioactive nucleus. But there is reason for optimism. We have a mechanism to produce the energy, and models might be made to reproduce the time scale. Models can suggest guidance. It is also interesting to consult information from topologically stable monopoles that condensed matter experiments have recently been able to produce, and subsequently destroy\cite{Ollikainen2017-ob}. \eit

\section{Consistency Checks} 

\label{sec:checks}
 
\subsection{Production Rates} 

The comprehensive monopole review of Mitsou\cite{Mitsou:2018hgt} includes more than 100 references with emphasis on experiments over the years, plus new LHC searches including MoEDAL. So far large detectors have set the most strict bounds. RICE\cite{Hogan:2008sx} and ANITA-II \cite{ANITA-II:2010jck} used radio detection to put upper limits on ultra-relativistic monopole fluxes of order $10^{-18}-10^{-19} \, cm^{-2}s^{-1}str^{-1}$ respectively. The Pierre Auger Observatory\cite{PierreAuger:2016imq}, set limits of order $ 10^{-15} - 10^{21} \, cm^{-2}s^{-1}str^{-1}$ depending on the boost factor. The current limits of ANTARES\cite{Boumaaza:2021ghi} and IceCube\cite{IceCube:2021eye} on ``slow'' monopoles with speed $0.5<\beta<0.995$ are claimed to be the most stringent. 

While these are very large detectors, most nations are larger. For example, the contiguous United States has an area of about $10^{7}\, km^{2}=10^{17}\, cm^{2}$. It is instrumented with about 350M channels, each capable of observing a volume of order $km^{3}$, which work on a duty cycle of order 30\%. The detectors are also very poorly distributed, of variable reliability, and prone to reject signals that do not match their triggers. Observing of order 100 events per year in the US corresponds to a flux of order $10^{-22}/cm^{-2}s^{-1}$ with an efficiency of order $10^{-8}$. In the sequel we will suggest ways efficiency could be approved.

Strict limits on the number of monopoles per nucleon of less than $10^{-29}$ have been set by moving about 1000 kg of material through superconducting loops\cite{PhysRevA.33.1183, Jeon:1995rf}. The material examined has included meteorites, schists, ferromanganese nodules, iron ores, and seawater. Ref. \cite{PhysRevD.8.698} is the last of a series of papers examining about 20 kg total material from the Moon. Refs.\cite{Kalbfleisch:2000iz,Kalbfleisch:2003yt,MoEDAL:2024wbc} searched material irradiated in particle accelerators. We found no record of studying vaporized or gaseous material. 
 
An upper limit on producing anything by lightning comes from the frequency of lighting strokes, which is of order $1/km^{2}/year$, with large regional variations. The electric field of a stroke spreads over a considerable area, often causing injuries to people who are not directly hit. Suppose a ``large'' electric field on the ground occurs over an area of $1 \, m^{2}$ to a depth of 0.1 m, a volume of 0.1 $m^{3}$, containing about $10^{30}$ nucleons. (The number of atoms in the air along the path of a stroke is much smaller.) Then the laboratory abundance limit, allowing  order 1-10 monopoles to be produced per lighting stroke, is too weak to be relevant. 
In fact, about 10M lighting strokes per year connect with the ground in the US. Then lighting has a chance to interact with about $10^{37}$ atoms per year. An abundance of $10^{-29}$ monopoles per nucleus leads to $10^{12}$monopoles per year multiplied by the branching ratio to make a monopole.

It is interesting that the production rate of ball lightning in the interior of airplanes in flight seems to be anomalously high. This may be a clue. Typical commercial airplanes are hit by lightning once per 3000 hours. Airplanes in flight are subject to precipitation charging to high voltages, need ``static wicks'' at strategic points to remove static electricity, and very regularly observe meter-scale electrostatic discharge on windshields (which pilots wrongly call St. Elmo's fire). Many civilian and military pilots and passengers have seen ball lighting appear on windshields and windows and wander through the volume of a plane. See Ref\cite{Lowke2012} for an event recorded by military personnel, and Ref\cite{jennisonBallLightning1969} for an event witnessed by a radio astronomer. The phenomenon is created or liberated in the rest frame of the plane. We found no reports of an object at rest in the atmosphere traversing the interior of a plane with a relative speed of hundreds of miles per hour.  As Lowke\cite{Lowke2012} observed, observations of ball lightning in the cockpits of aircraft may have decreased. The cause may be increased use of conducting layers added to the surface for electric heating. 

\subsection{Radiation Signal} 

\label{sec:ionizing}

Many researchers have observed that a significant amount of power in ionizing or higher energy radiation might lead to radiation burns or poisoning of observers. A few rare cases match this very well, but they are exceptions. In most cases humans seem to be unaffected by seeing ball lightning at distances of a few meters and more. 

One $gray=1 \, J/kg=10^{2} \, rad$ is a physical unit of absorbed radiation. The unit of equivalent absorbed dose is the {\it sievert (Sv)}=J/kg$\times W_{R}$, where $W_{R}$ is a factor taking into account the biological effectiveness of different particles. 1 sievert (Si)=100 {\it rem} is a dose that usually makes people very sick or might kill them. One {\it Curie (Ci) } =$3.7 \times 10^{10}$ decays per second is a unit of activity ($A$). A health physics rule of thumb estimates the equivalent dose rate $\dot D$ from a point source of {\it MeV-scale} beta particles at a distance $L$ as  
\bn \dot D = {27 A \ov Ci}{1 \over  (L/m)^{2}}{rem \ov hr}. \ea If MeV-scale particles are produced with $10^{3}$W of power there are $10^{16}$ produced per second. With $A \sim 10^{6}$ and $L \sim 1m$ the equivalent dose rate of MeV betas is 75 Sv/s, which is surely deadly. (Yet the majority of events in three dimensional space are observed at the maximum possible distance. The $1/L^{2}$ dependence reduces the estimate by a factor of 10,000 at $L \sim$ 100 m.)

The range in air of 1 MeV electrons is about 1000 cm, and incompatible with ball lightning. If pairs are produced at threshold their energy will be much lower. For energies $0.0< E< 0.5$ MeV the range $R$ of electrons and positrons in air $R \sim 800 (E/MeV)^{1.6} cm$, with relative errors of 20\% or less. (Positron annihilation is a small effect in this regime compared to ionization and Compton scattering.) A beta particle with energy of 0.1 MeV has a range in air of 16 cm, which is a reasonable scale for ball lightning. This is yet another interesting numerical fact. 

It is very important that the effects of beta radiation depend strongly on energy. Beta particles with $E<0.070 keV$ cannot penetrate skin, and pose no risk to health. Indeed many beta emitting isotopes such as $^{3}H \, (0.02 \,MeV), \, ^{14}C \, (0.15 \,MeV, ^{129}I\, (0.15 MeV)$ are not treated as radioisotopes that require protection from external exposure. They are both safe, and too hard to detect, so they are ignored \cite{HaraKang}. Their gamma-ray equivalent factors are often listed as zero on health physics data sheets. (Health consequences from ingestion and long term incorporation into the body are a different matter.) As a specific example\cite{sheet}, 1 Ci of $^{89}Sr$ (a beta emitter with energy 585 keV) produces a dose rate of 0.00041 rem/hr. An activity of $10^{16}/s$ makes a dose rate of $10^{-3} Sv s^{-1}$, which is negligible.

\subsection{The 511 keV Signal}

In full view of large uncertainties, the size scale and radiation characteristics of an agent decaying to $e^{+}e^{-}$ pairs close to threshold are consistent enough with the phenomenology of ball lightning. We also need to estimate the rate of annihilation, which would produce a distinctive signal of 0.511 gamma rays. If $10^{16}$ pairs are spread over a 10 cm ball, the pair number density is order $10^{13}cm^{-3}$, which is much smaller than number density of air. The air around the object will be hot, say at temperature $T$, with the number density scaling like $1/T$ and thermal velocity scaling like $\sqrt{T}$ at constant pressure. An upper bound on a annihilation rate is approximately \ba <\sigma n v> \sim 2 \times 10^{-3}s^{-1}({\sigma \over 10^{24}cm^{2}})({<v_{STP}> \over 5 \times 10^{4}cm/s})({100 keV \over \sqrt{T}  }). \nn \ea This is an upper bound because hot electrons are faster than nuclei, and rapidly depart from an ionized region, creating a so-called Langmuir layer of charge separation. If $10^{16}$ pairs are produced per second, at most $10^{-3}$ of them can annihilate to make $10^{13}$ gamma rays per second. We divide the previous estimate of radiation dose by 1000, yielding 10 rem/s at a distance of 1m. According to Ref\cite{website}, Title 10, Part 20, of the Code of Federal Regulations establishes the dose limits for radiation workers\footnote{We notice the Code lets radiation workers have a higher dose limit than other humans. We did not find out why. } The annual total for the whole body is 5 rem. It appears that a brief encounter with ball lightning more or less doubles a radiation worker's annual exposure, up to certain assumptions. The calculation is not stable, being very sensitive to the electron density in the vicinity of the object, and directly proportional to the power, which makes two uncertain factors. The $1/L^{2}$ decrease of radiation flux with distance is also a significant effect. In any event this ``numerical coincidence'' is remarkable. In initial research for the paper we thought radiation would kill the whole idea. 

The idea that ball lightning might produce 511 keV gamma rays is not new. As reviewed by Ref. \cite{Shmatov2005-zt}, many plasma ball models \cite{Shmatov2003-in,Altschuler1970-ra,Ashby1971-dk,Dijkhuis1980-ni} predict gamma rays, while (we reiterate) being incapable of explaining all the data, including the source of power or passage through solid obstacles. Experimental detection was investigated by Ashby and Whitehead in 1971, who found events in a nighttime sky survey. Ordinary astrophysical backgrounds explain those events. See Ref.\cite{Adriani2011-yl} for satellite measurements of geomagnetically trapped cosmic ray antiprotons.) There would seem to be almost no possible background in Nature to produce a continuous flux of 511 keV radiation over a time scale of many seconds. (An anti-matter meteorite would do it, which was what Ashby and Whitehead had imagined.)

In fact there is a large literature on the production of gamma rays by thunderstorms. Dwyer {\it et al} Ref. \cite{dwyerPositronCloudsThunderstorms2015} is cited for observing 511 keV spectral peaks in thunderstorms from positron annihilation. Ref. \cite{kochkinFlightObservationPositron2018} observed the 511 keV peaks, along with correlations between gamma ray bursts of 1 s time scale and electric field pulses. Recently Ref. \cite{ostgaardFlickeringGammarayFlashes2024} has observed ``flickering gamma-ray flashes'' on time scales of 10-1000 $\mu s$. All of these observations come from aircraft-borne instruments. A different class of events called ``gamma ray glows'' (first observed by Parks {\it et al}\cite{Parks1981} have energies up to tens of MeV and durations ranging from seconds to several minutes\cite{wadaCatalogGammarayGlows2021}. Wada {\it et al}\cite{wadaCatalogGammarayGlows2021} lists the spectra and count rate histories of a catalog of 70 events from the ground based GROWTH experiment in Japan. The averaged energy spectra of events, unfolded to account for detector efficiency, (paper Figure 12) shows an apparent 0.511 keV peak. The peak is not visible in most event spectra. 

The mechanism proposed for generating positrons in thunderstorms is photonuclear reactions involving Nitrogen. Photonuclear cross sections of multi-MeV electrons and positrons are much smaller than bremsstrahlung and pair production cross sections. The photonuclear mechanism needs enormous numbers of photons with energies above about 20 MeV to be important. 

A very interesting lightning-associated event {\it with a 511 keV peak persisting for about a minute} was observed by Enoto {\it et al}\cite{enoto2017,yuasaThundercloudProjectExploring2020}. The detectors monitored winter thunderstorms on the coast of Japan, which occur unusually close to the ground. 

\section{Discussion} 

\label{sec:discussion}

There may be 10,000 or more PhD physicists energetically searching for physics beyond the Standard Model. The absence of mobilization to determine the nature of ball lighting is both strange and understandable. Most physicists have never heard of ball lighting. Those who heard have assumed it must have some conventional explanation within chemistry or plasma physics. (If something is assumed to be conventional, few bother to check it.) The most important message is that there is no conventional explanation in the Standard Model. We can only begin to imagine the intensity that teams of experienced experimental physicists could bring to the subject. 

Here are a few suggestions: \bit \im Physicists should not be idle, but leaders organizing the public, if only to get the public aware that {\it searches for new physics are legitimate and exciting.} That could make a significant contribution if nothing but limits were established. And what if an informed public becomes engaged? \im Direct searches for monopole abundance in matter strangely stopped after examining about 1000 kg, of what may have been the wrong materials. The association of ball lightning with glass may be a clue. There is plenty of glass to recycle, and can imagine an automated process searching hundreds on tons of material. (Recall that when proton decay attracted serious consideration, the scale of detectors expanded by something like $10^{4}$). The frequency of ball lightning occurring inside aircraft may be a clue. \im Currently the triggers and data analysis for facilities with very large detection volumes are wrong to contribute much. If a non-relativistic event with thousands of watts triggered a large array, it might be rejected as a transient equipment malfunction, or a meteor, or an uninteresting atmospheric glitch. It is noteworthy that some cosmic ray facilities, e. g. the Pierre Auger observatory\cite{inproceedings} and the Telescope Array\cite{abbasiGammaRayShowers2018,AbbasiTelescopeCoincidence} do participate in lightning studies. While detectors with the largest effective volumes have not set useful limits, we can imagine that more useful limits could be found with appropriate search strategies. \im According to a US Department of Labor report \cite{LaborStatistics}, the number of security cameras in the US grew from 47 $\ra$ 70 $\ra$ 85 million in the three years of 2018-2021. There are well over 100 million cameras now. Many of them are outdoors, monitoring traffic and security over large volumes day and night. (Stephan\cite{stephan2024balllightningmagneticmonopoles} has also mentioned this.) The exponential rise of artificial intelligence is already blooming into a multi-billion dollar video monitoring industry. Adding an algorithm trained to detect ball lightning is clearly feasible. Projects to coordinate such activity are perfectly suited to the ambitious large-scale management skills of high energy, nuclear, and cosmic ray experimental physicists. \eit

\subsection{Organization Is Needed} 

We conclude with a reminder that many ambitious searches for new physics have been proposed and funded on the basis of a projected rate of 100 or less events per year. There has never been a significant coordinated effort by physicists to explore ball lightning, and the public is not aware of its importance. The anecdotal event rate must be a lower bound on the actual rate. Exploring physics beyond the Standard Model is always challenging. It is delightful to realize that exploring ball lightning does not need the mammoth resources of a new particle accelerator.

\section{Acknowledgements} 
I thank Dave Besson, David Fryberger, Ian Lewis, Kim Milton, Steven Prohira, Richard Sonnenfeld, Karl Stephan, and  Daniel Tapia Takaki for discussions and comments.

\bibliography{bibby}{}

\begin{thebibliography}{10}

\bibitem{Uman1969-qj}
Martin~A Uman.
\newblock {\em Lightning}.
\newblock McGraw-Hill advanced physics monograph series. McGraw-Hill, New York,
  NY, February 1969.

\bibitem{Barry1980}
James~Dale Barry.
\newblock {\em Ball Lightning}, pages 33--43.
\newblock Springer US, Boston, MA, 1980.

\bibitem{Stenhoff2010-by}
Mark Stenhoff.
\newblock {\em Ball lightning}.
\newblock Springer, New York, NY, 1 edition, November 2010.

\bibitem{Boerner2019-eh}
Herbert Boerner.
\newblock {\em Ball lightning}.
\newblock Springer Nature, Cham, Switzerland, 1 edition, July 2019.

\bibitem{Smirnov1993-qu}
B~M Smirnov.
\newblock Physics of ball lightning.
\newblock {\em Phys. Rep.}, 224(4):151--236, March 1993.

\bibitem{RakovUman2003}
Vladimir~A. Rakov and Martin~A. Uman.
\newblock {\em Lightning: Physics and Effects}, chapter 20, Ball lightning,
  bead lightning, and other unusual discharges, pages 656--674.
\newblock Cambridge University Press, 2003.

\bibitem{hgss-12-43-2021}
A.~G. Keul.
\newblock A brief history of ball lightning observations by scientists and
  trained professionals.
\newblock {\em History of Geo- and Space Sciences}, 12(1):43--56, 2021.

\bibitem{Wu2016-ju}
H-C Wu.
\newblock Relativistic-microwave theory of ball lightning.
\newblock {\em Sci. Rep.}, 6(1):28263, June 2016.

\bibitem{Shmatov2019-gt}
Mikhail~L Shmatov and Karl~D Stephan.
\newblock Advances in ball lightning research.
\newblock {\em J. Atmos. Sol. Terr. Phys.}, 195(105115):105115, November 2019.

\bibitem{PhysRev.135.A390}
David Finkelstein and Julio Rubinstein.
\newblock Ball lightning.
\newblock {\em Phys. Rev.}, 135:A390--A396, Jul 1964.

\bibitem{Shmatov2003-in}
M~L Shmatov.
\newblock New model and estimation of the danger of ball lightning.
\newblock {\em J. Plasma Phys.}, 69(6):507--527, December 2003.

\bibitem{Lowke2012}
J.~J. Lowke, D.~Smith, K.~E. Nelson, R.~W. Crompton, and A.~B. Murphy.
\newblock Birth of ball lightning.
\newblock {\em Journal of Geophysical Research: Atmospheres}, 117(D19), 2012.

\bibitem{Morrow2017-aw}
R~Morrow.
\newblock Ball lightning dynamics and stability at moderate ion densities.
\newblock {\em J. Phys. D Appl. Phys.}, 50(39):395201, October 2017.

\bibitem{Abrahamson2002}
J.~Abrahamson and J.~Diniss.
\newblock {Ball lightning caused by oxidation of nanoparticle networks from
  normal lightning strikes on soil}.
\newblock {\em Nature}, 403:519--521, 2002.

\bibitem{stephan2024balllightningmagneticmonopoles}
Karl~D. Stephan.
\newblock Could ball lightning be magnetic monopoles?
\newblock https://arxiv.org/abs/2408.10289, 2024.

\bibitem{Korshunov:1990gf}
V.~K. Korshunov.
\newblock {Drift motion of the magnetic monopole of Polyakov-'t Hooft in the
  air and the 'ball lightning' phenomenon}.
\newblock {\em Mod. Phys. Lett. A}, 5:1629--1631, 1990.

\bibitem{Rubakov:1981rg}
V.~A. Rubakov.
\newblock {Superheavy Magnetic Monopoles and Proton Decay}.
\newblock {\em JETP Lett.}, 33:644--646, 1981.

\bibitem{Rubakov:1982fp}
V.~A. Rubakov.
\newblock {Adler-Bell-Jackiw Anomaly and Fermion Number Breaking in the
  Presence of a Magnetic Monopole}.
\newblock {\em Nucl. Phys. B}, 203:311--348, 1982.

\bibitem{Callan:1982ac}
Curtis~G. Callan, Jr.
\newblock {Monopole Catalysis of Baryon Decay}.
\newblock {\em Nucl. Phys. B}, 212:391--400, 1983.

\bibitem{Callan:1982au}
Curtis~G. Callan, Jr.
\newblock {Dyon-Fermion Dynamics}.
\newblock {\em Phys. Rev. D}, 26:2058--2068, 1982.

\bibitem{Dirac:1931kp}
Paul Adrien~Maurice Dirac.
\newblock {Quantised Singularities in the Electromagnetic Field}.
\newblock {\em Proc. Roy. Soc. Lond. A}, 133(821):60--72, 1931.

\bibitem{tHooft:1974kcl}
Gerard 't~Hooft.
\newblock {Magnetic Monopoles in Unified Gauge Theories}.
\newblock {\em Nucl. Phys. B}, 79:276--284, 1974.

\bibitem{Polyakov:1974ek}
Alexander~M. Polyakov.
\newblock {Particle Spectrum in Quantum Field Theory}.
\newblock {\em JETP Lett.}, 20:194--195, 1974.

\bibitem{Kazama:1976fm}
Yoichi Kazama, Chen~Ning Yang, and Alfred~S. Goldhaber.
\newblock {Scattering of a Dirac Particle with Charge Ze by a Fixed Magnetic
  Monopole}.
\newblock {\em Phys. Rev. D}, 15:2287--2299, 1977.

\bibitem{ParticleDataGroup:2024cfk}
S.~Navas et~al.
\newblock {Review of particle physics}.
\newblock {\em Phys. Rev. D}, 110(3):030001, 2024.

\bibitem{10.1063/PT.3.3328}
Arttu Rajantie.
\newblock The search for magnetic monopoles.
\newblock {\em Physics Today}, 69(10):40--46, October 2016.

\bibitem{RajanteMonopoles}
Arttu Rajantie.
\newblock Introduction to magnetic monopoles.
\newblock {\em Contemporary Physics}, 53(3):195--211, 2012.

\bibitem{Milton:2006cp}
Kimball~A. Milton.
\newblock {Theoretical and experimental status of magnetic monopoles}.
\newblock {\em Rept. Prog. Phys.}, 69:1637--1712, 2006.

\bibitem{giacomelliMagneticMonopoleBibliography2000b}
G.~Giacomelli, M.~Giorgini, T.~Lari, M.~Ouchrif, L.~Patrizii, V.~Popa,
  P.~Spada, and V.~Togo.
\newblock Magnetic monopole bibliography, May 2000.

\bibitem{Giacomelli:1984gq}
G.~Giacomelli.
\newblock {Magnetic Monopoles}.
\newblock {\em Riv. Nuovo Cim.}, 7N12:1--111, 1984.

\bibitem{Stephan2016-el}
Karl~D Stephan, Rozlyn Krajcik, and Rolf~J Martin.
\newblock Fluorescence caused by ionizing radiation from ball lightning:
  Observation and quantitative analysis.
\newblock {\em J. Atmos. Sol. Terr. Phys.}, 148:32--38, October 2016.

\bibitem{Fryberger:1994vv}
David Fryberger.
\newblock {A Model for ball lightning}.
\newblock In {\em {1st International Workshop on the Unidentified Atmospheric
  Light Phenomena in Hessdalen}}, 10 1994.

\bibitem{Lipkin:1983gp}
Harry~J. Lipkin.
\newblock {Effects of Magnetic Monopoles on Nuclear Wave Functions and Possible
  Catalysis of Nuclear beta Decay and Spontaneous Fission}.
\newblock {\em Phys. Lett. B}, 133:347, 1983.

\bibitem{Sakurai2020-ji}
J~J Sakurai and Jim Napolitano.
\newblock {\em Modern quantum mechanics}.
\newblock Cambridge University Press (Virtual Publishing), Cambridge, England,
  3 edition, October 2020.

\bibitem{Sen:1983yq}
Ashoke Sen.
\newblock {Conservation Laws in the Monopole Induced Baryon Number Violating
  Processes}.
\newblock {\em Phys. Rev. D}, 28:876, 1983.

\bibitem{Workman:2022ynf}
R.~L. Workman and Others.
\newblock {Review of Particle Physics}.
\newblock {\em PTEP}, 2022:083C01, 2022.

\bibitem{Backovic:2009rw}
Mihailo Backovic and John~P. Ralston.
\newblock {Limits on Threshold and 'Sommerfeld' Enhancements in Dark Matter
  Annihilation}.
\newblock {\em Phys. Rev. D}, 81:056002, 2010.

\bibitem{miltonCoulombResummationMonopole2008c}
K.~A. Milton.
\newblock Coulomb {{Resummation}} and {{Monopole Masses}}, February 2008.

\bibitem{cecchiniEnergyLossesMagnetic2016}
S.~Cecchini, L.~Patrizii, Z.~Sahnoun, G.~Sirri, and V.~Togo.
\newblock Energy {{Losses}} of {{Magnetic Monopoles}} in {{Aluminum}}, {{Iron}}
  and {{Copper}}, June 2016.

\bibitem{Ohtsuki1991-pg}
Y~H Ohtsuki and H~Ofuruton.
\newblock Plasma fireballs formed by microwave interference in air.
\newblock {\em Nature}, 350(6314):139--141, March 1991.

\bibitem{anthonyExperimentalStudiesLight2009}
P.L. Anthony, J.R. Delayen, D.~Fryberger, W.S. Goree, J.~Mammosser, Z.M.
  Szalata, and J.G. Weisend.
\newblock Experimental studies of light emission phenomena in superconducting
  {{RF}} cavities.
\newblock {\em Nuclear Instruments and Methods in Physics Research Section A:
  Accelerators, Spectrometers, Detectors and Associated Equipment},
  612(1):1--45, December 2009.

\bibitem{PhysRev.83.899}
W.~V.~R. Malkus.
\newblock The interaction of the dirac magnetic monopole with matter.
\newblock {\em Phys. Rev.}, 83:899--905, Sep 1951.

\bibitem{sivers1970possible}
Dennis Sivers.
\newblock Possible binding of a magnetic monopole to a particle with electric
  charge and a magnetic dipole moment.
\newblock {\em Physical Review D}, 2(9):2048, 1970.

\bibitem{Bracci1984-zy}
L~Bracci and G~Fiorentini.
\newblock Interactions of magnetic monopoles with nuclei and atoms: Formation
  of bound states and phenomenological consequences.
\newblock {\em Nucl. Phys. B.}, 232(2):236--262, February 1984.

\bibitem{Olaussen:1984xb}
K.~Olaussen and R.~Sollie.
\newblock {Form-Factor Effects On Nucleus Magnetic Monopole Binding}.
\newblock {\em Nucl. Phys. B}, 255:465--479, 1985.

\bibitem{Kazama:1976sr}
Yoichi Kazama and Chen~Ning Yang.
\newblock {Existence of Bound States for a Charged Spin 1/2 Particle with an
  Extra Magnetic Moment in the Field of a Fixed Magnetic Monopole}.
\newblock {\em Phys. Rev. D}, 15:2300, 1977.

\bibitem{Jackiw:1975fn}
R.~Jackiw and C.~Rebbi.
\newblock {Solitons with Fermion Number 1/2}.
\newblock {\em Phys. Rev. D}, 13:3398--3409, 1976.

\bibitem{Callan:1982ah}
Curtis~G. Callan, Jr.
\newblock {Disappearing Dyons}.
\newblock {\em Phys. Rev. D}, 25:2141, 1982.

\bibitem{Balachandran:1983fh}
A.~P. Balachandran and J.~Schechter.
\newblock {Monopole Induced Proton Disintegration}.
\newblock {\em Phys. Rev. D}, 29:1184, 1984.

\bibitem{Weinberg:1998aw}
Erick~J. Weinberg.
\newblock {Massive monopoles and massless monopole clouds}.
\newblock In {\em {International Workshop on Mathematical and Physical Aspects
  of Nonlinear Field Theories}}, 2 1998.

\bibitem{Gervalle2022-kk}
Romain Gervalle and Mikhail~S Volkov.
\newblock Electroweak monopoles and their stability.
\newblock {\em Nucl. Phys. B.}, 984(115937):115937, November 2022.

\bibitem{Steinhardt:1981mm}
Paul~Joseph Steinhardt.
\newblock {Monopole Dissociation in the Early Universe}.
\newblock {\em Phys. Rev. D}, 24:842, 1981.

\bibitem{Callan:1983nx}
Curtis~G. Callan, Jr. and Edward Witten.
\newblock {Monopole Catalysis of Skyrmion Decay}.
\newblock {\em Nucl. Phys. B}, 239:161--176, 1984.

\bibitem{Bais:2002ae}
F.~A. Bais and J.~Striet.
\newblock {On a core instability of 't Hooft-Polyakov monopoles}.
\newblock {\em Phys. Lett. B}, 540:319--323, 2002.

\bibitem{striet_more_2003}
J.~Striet and F.~A. Bais.
\newblock More on core instabilities of magnetic monopoles.
\newblock {\em Journal of High Energy Physics}, 2003(06):022--022, June 2003.
\newblock arXiv:hep-th/0304189.

\bibitem{Cho:1996qd}
Y.~M. Cho and D.~Maison.
\newblock {Monopoles in Weinberg-Salam model}.
\newblock {\em Phys. Lett. B}, 391:360--365, 1997.

\bibitem{Heisenberg:1936nmg}
W.~Heisenberg and H.~Euler.
\newblock {Consequences of Dirac's theory of positrons}.
\newblock {\em Z. Phys.}, 98(11-12):714--732, 1936.

\bibitem{Schwinger:1951nm}
Julian~S. Schwinger.
\newblock {On gauge invariance and vacuum polarization}.
\newblock {\em Phys. Rev.}, 82:664--679, 1951.

\bibitem{doi:10.1142/9789812775344_0014}
Gerald~V. Dunne.
\newblock {\em Heisenberg-Euler Effective Lagrangians: Basics and Extensions},
  pages 445--522.
\newblock World Scientific, 2005.

\bibitem{Ollikainen2017-ob}
T~Ollikainen, K~Tiurev, A~Blinova, W~Lee, D~S Hall, and M~M{\"o}tt{\"o}nen.
\newblock Experimental realization of a dirac monopole through the decay of an
  isolated monopole.
\newblock {\em Phys. Rev. X.}, 7(2), May 2017.

\bibitem{Mitsou:2018hgt}
Vasiliki~A. Mitsou.
\newblock {The quest for magnetic monopoles \textendash{} past, present and
  future}.
\newblock {\em PoS}, CORFU2017:188, 2018.

\bibitem{Hogan:2008sx}
D.~P. Hogan, D.~Z. Besson, J.~P. Ralston, I.~Kravchenko, and D.~Seckel.
\newblock {Relativistic Magnetic Monopole Flux Constraints from RICE}.
\newblock {\em Phys. Rev. D}, 78:075031, 2008.

\bibitem{ANITA-II:2010jck}
M.~Detrixhe et~al.
\newblock {Ultra-Relativistic Magnetic Monopole Search with the ANITA-II
  Balloon-borne Radio Interferometer}.
\newblock {\em Phys. Rev. D}, 83:023513, 2011.

\bibitem{PierreAuger:2016imq}
Alexander Aab et~al.
\newblock {Search for ultrarelativistic magnetic monopoles with the Pierre
  Auger Observatory}.
\newblock {\em Phys. Rev. D}, 94(8):082002, 2016.

\bibitem{Boumaaza:2021ghi}
Jihad Boumaaza, J.~Brunner, A.~Moussa, and Y.~Tayalati.
\newblock {Search for relativistic Magnetic Monopoles with ten years of the
  ANTARES detector data}.
\newblock {\em PoS}, ICRC2021:1127, 2021.

\bibitem{IceCube:2021eye}
R.~Abbasi et~al.
\newblock {Search for Relativistic Magnetic Monopoles with Eight Years of
  IceCube Data}.
\newblock {\em Phys. Rev. Lett.}, 128(5):051101, 2022.

\bibitem{PhysRevA.33.1183}
Joseph~M. Kovalik and Joseph~L. Kirschvink.
\newblock New superconducting-quantum-interference-device-based constraints on
  the abundance of magnetic monopoles trapped in matter: An investigation of
  deeply buried rocks.
\newblock {\em Phys. Rev. A}, 33:1183--1187, Feb 1986.

\bibitem{Jeon:1995rf}
Hunmoo Jeon and Michael~J. Longo.
\newblock {Search for magnetic monopoles trapped in matter}.
\newblock {\em Phys. Rev. Lett.}, 75:1443--1446, 1995.
\newblock [Erratum: Phys.Rev.Lett. 76, 159 (1996)].

\bibitem{PhysRevD.8.698}
Ronald~R. Ross, Philippe~H. Eberhard, Luis~W. Alvarez, and Robert~D. Watt.
\newblock Search for magnetic monopoles in lunar material using an
  electromagnetic detector.
\newblock {\em Phys. Rev. D}, 8:698--702, Aug 1973.

\bibitem{Kalbfleisch:2000iz}
G.~R. Kalbfleisch, K.~A. Milton, Michael~G. Strauss, L.~P. Gamberg, E.~H.
  Smith, and W.~Luo.
\newblock {Improved experimental limits on the production of magnetic
  monopoles}.
\newblock {\em Phys. Rev. Lett.}, 85:5292--5295, 2000.

\bibitem{Kalbfleisch:2003yt}
G.~R. Kalbfleisch, W.~Luo, K.~A. Milton, E.~H. Smith, and Michael~G. Strauss.
\newblock {Limits on production of magnetic monopoles utilizing samples from
  the D0 and CDF detectors at the Tevatron}.
\newblock {\em Phys. Rev. D}, 69:052002, 2004.

\bibitem{MoEDAL:2024wbc}
B.~Acharya et~al.
\newblock {MoEDAL Search in the CMS Beam Pipe for Magnetic Monopoles Produced
  via the Schwinger Effect}.
\newblock {\em Phys. Rev. Lett.}, 133(7):071803, 2024.

\bibitem{jennisonBallLightning1969}
R.~C. Jennison.
\newblock Ball {{Lightning}}.
\newblock {\em Nature}, 224(5222):895--895, November 1969.

\bibitem{HaraKang}
Hara Kang, Sujung Min, Bumkyung Seo, Changhyun Roh, Sangbum Hong, and Jae~Hak
  Cheong.
\newblock Low energy beta emitter measurement: A review.
\newblock {\em Chemosensors}, 8:106, 2020.

\bibitem{sheet}
Stanford university health~data sheets.
\newblock \url{https://ehs.stanford.edu/wp-content/uploads/Sr-89-RSDS.pdf}.

\bibitem{website}
Stanford University~Environmental Health and ''3.2 Maximum Permissible
  Occupational~Doses'' Safety.
\newblock
  \url{https://ehs.stanford.edu/manual/radiation-protection-guidance-hospital-staff/maximum-permissible-occupational-doses#:~:text=Title%2010%2C%20Part%2020%2C%20of,whole%20body%20is%205%2C000%20mrem.}

\bibitem{Shmatov2005-zt}
M~L Shmatov.
\newblock Expected spectrum of high-energy photons from ball lightning.
\newblock {\em J. Plasma Phys.}, 72(02):277, December 2005.

\bibitem{Altschuler1970-ra}
M~D Altschuler, L~L House, and E~Hildner.
\newblock Is ball lightning a nuclear phenomenon ?
\newblock {\em Nature}, 228(5271):545--547, November 1970.

\bibitem{Ashby1971-dk}
D~E T~F Ashby and C~Whitehead.
\newblock Is ball lightning caused by antimatter meteorites?
\newblock {\em Nature}, 230(5290):180--182, March 1971.

\bibitem{Dijkhuis1980-ni}
G~C Dijkhuis.
\newblock A model for ball lightning.
\newblock {\em Nature}, 284(5752):150--151, March 1980.

\bibitem{Adriani2011-yl}
O.~et~al Adriani.
\newblock The discovery of geomagnetically trapped cosmic-ray antiprotons.
\newblock {\em Astrophys. J. Lett.}, 737(2):L29, August 2011.

\bibitem{dwyerPositronCloudsThunderstorms2015}
Joseph~R. Dwyer, David~M. Smith, Bryna~J. Hazelton, Brian~W. Grefenstette,
  Nicole~A. Kelley, Alexander~W. Lowell, Meagan~M. Schaal, and Hamid~K.
  Rassoul.
\newblock Positron clouds within thunderstorms.
\newblock {\em Journal of Plasma Physics}, 81(4):475810405, August 2015.

\bibitem{kochkinFlightObservationPositron2018}
P.~Kochkin, D.~Sarria, C.~Skeie, A.~P.~J. {van Deursen}, A.~I. {de Boer},
  M.~Bardet, C.~Allasia, F.~Flourens, and N.~{\O}stgaard.
\newblock In-{{Flight Observation}} of {{Positron Annihilation}} by {{ILDAS}}.
\newblock {\em Journal of Geophysical Research: Atmospheres},
  123(15):8074--8090, 2018.

\bibitem{ostgaardFlickeringGammarayFlashes2024}
N.~{\O}stgaard, A.~Mezentsev, M.~Marisaldi, J.~E. Grove, M.~Quick,
  H.~Christian, S.~Cummer, M.~Pazos, Y.~Pu, M.~Stanley, D.~Sarria, T.~Lang,
  C.~Schultz, R.~Blakeslee, I.~Adams, R.~Kroodsma, G.~Heymsfield, N.~Lehtinen,
  K.~Ullaland, S.~Yang, B.~Hasan Qureshi, J.~S{\o}ndergaard, B.~Husa,
  D.~Walker, D.~Shy, M.~Bateman, P.~Bitzer, M.~Fullekrug, M.~Cohen,
  J.~Montanya, C.~Younes, O.~Van Der~Velde, P.~Krehbiel, J.~A. Roncancio, J.~A.
  Lopez, M.~Urbani, A.~Santos, and D.~Mach.
\newblock Flickering gamma-ray flashes, the missing link between gamma glows
  and {{TGFs}}.
\newblock {\em Nature}, 634(8032):53--56, October 2024.

\bibitem{Parks1981}
G.~K. Parks, B.~H. Mauk, R.~Spiger, and J.~Chin.
\newblock X-ray enhancements detected during thunderstorm and lightning
  activities.
\newblock {\em Geophysical Research Letters}, 8(11):1176--1179, 1981.

\bibitem{wadaCatalogGammarayGlows2021}
Y.~Wada, T.~Matsumoto, T.~Enoto, K.~Nakazawa, T.~Yuasa, Y.~Furuta, D.~Yonetoku,
  T.~Sawano, G.~Okada, H.~Nanto, S.~Hisadomi, Y.~Tsuji, G.~S. Diniz,
  K.~Makishima, and H.~Tsuchiya.
\newblock Catalog of gamma-ray glows during four winter seasons in {{Japan}}.
\newblock {\em Physical Review Research}, 3(4):043117, November 2021.

\bibitem{enoto2017}
Teruaki Enoto, Yuuki Wada, Yoshihiro Furuta, Kazuhiro Nakazawa, Takayuki Yuasa,
  Kazufumi Okuda, Kazuo Makishima, Mitsuteru Sato, Yousuke Sato, Toshio Nakano,
  Daigo Umemoto, and Harufumi Tsuchiya.
\newblock Photonuclear reactions triggered by lightning discharge.
\newblock {\em Nature}, 551(7681):481--484, November 2017.

\bibitem{yuasaThundercloudProjectExploring2020}
Takayuki Yuasa, Yuuki Wada, Teruaki Enoto, Yoshihiro Furuta, Harufumi Tsuchiya,
  Shohei Hisadomi, Yuna Tsuji, Kazufumi Okuda, Takahiro Matsumoto, Kazuhiro
  Nakazawa, Kazuo Makishima, Shoko Miyake, and Yuko Ikkatai.
\newblock Thundercloud {{Project}}: {{Exploring}} high-energy phenomena in
  thundercloud and lightning.
\newblock {\em Progress of Theoretical and Experimental Physics},
  2020(10):103H01, October 2020.

\bibitem{inproceedings}
Julian Rautenberg.
\newblock Lightning detection at the pierre auger observatory.
\newblock page 678, 08 2016.

\bibitem{abbasiGammaRayShowers2018}
R.~U. Abbasi, T.~Abu-Zayyad, M.~Allen, E.~Barcikowski, J.~W. Belz, D.~R.
  Bergman, S.~A. Blake, M.~Byrne, R.~Cady, B.G. Cheon, J.~Chiba, M.~Chikawa,
  T.~Fujii, M.~Fukushima, G.~Furlich, T.~Goto, W.~Hanlon, Y.~Hayashi,
  N.~Hayashida, K.~Hibino, K.~Honda, D.~Ikeda, N.~Inoue, T.~Ishii, H.~Ito,
  D.~Ivanov, S.~Jeong, C.~C.~H. Jui, K.~Kadota, F.~Kakimoto, O.~Kalashev,
  K.~Kasahara, H.~Kawai, S.~Kawakami, K.~Kawata, E.~Kido, H.~B. Kim, J.~H. Kim,
  J.~H. Kim, S.~S. Kishigami, P.~R. Krehbiel, V.~Kuzmin, Y.~J. Kwon, J.~Lan,
  R.~LeVon, J.~P. Lundquist, K.~Machida, K.~Martens, T.~Matuyama, J.~N.
  Matthews, M.~Minamino, K.~Mukai, I.~Myers, S.~Nagataki, R.~Nakamura,
  T.~Nakamura, T.~Nonaka, S.~Ogio, M.~Ohnishi, H.~Ohoka, K.~Oki, T.~Okuda,
  M.~Ono, R.~Onogi, A.~Oshima, S.~Ozawa, I.~H. Park, M.~S. Pshirkov,
  J.~Remington, W.~Rison, D.~Rodeheffer, D.~C. Rodriguez, G.~Rubtsov, D.~Ryu,
  H.~Sagawa, K.~Saito, N.~Sakaki, N.~Sakurai, T.~Seki, K.~Sekino, P.D. Shah,
  F.~Shibata, T.~Shibata, H.~Shimodaira, B.~K. Shin, H.~S. Shin, J.~D. Smith,
  P.~Sokolsky, R.~W. Springer, B.~T. Stokes, T.~A. Stroman, H.~Takai,
  M.~Takeda, R.~Takeishi, A.~Taketa, M.~Takita, Y.~Tameda, H.~Tanaka,
  K.~Tanaka, M.~Tanaka, R.~J. Thomas, S.~B. Thomas, G.~B. Thomson, P.~Tinyakov,
  I.~Tkachev, H.~Tokuno, T.~Tomida, S.~Troitsky, Y.~Tsunesada, Y.~Uchihori,
  S.~Udo, F.~Urban, G.~Vasiloff, T.~Wong, M.~Yamamoto, R.~Yamane, H.~Yamaoka,
  K.~Yamazaki, J.~Yang, K.~Yashiro, Y.~Yoneda, S.~Yoshida, H.~Yoshii, and
  Z.~Zundel.
\newblock Gamma {{Ray Showers Observed}} at {{Ground Level}} in {{Coincidence
  With Downward Lightning Leaders}}.
\newblock {\em Journal of Geophysical Research: Atmospheres},
  123(13):6864--6879, July 2018.

\bibitem{AbbasiTelescopeCoincidence}
R.~U. Abbasi, T.~Abu-Zayyad, M.~Allen, E.~Barcikowski, J.~W. Belz, D.~R.
  Bergman, S.~A. Blake, M.~Byrne, R.~Cady, B.G. Cheon, J.~Chiba, M.~Chikawa,
  T.~Fujii, M.~Fukushima, G.~Furlich, T.~Goto, W.~Hanlon, Y.~Hayashi,
  N.~Hayashida, K.~Hibino, K.~Honda, D.~Ikeda, N.~Inoue, T.~Ishii, H.~Ito,
  D.~Ivanov, S.~Jeong, C.~C.~H. Jui, K.~Kadota, F.~Kakimoto, O.~Kalashev,
  K.~Kasahara, H.~Kawai, S.~Kawakami, K.~Kawata, E.~Kido, H.~B. Kim, J.~H. Kim,
  J.~H. Kim, S.~S. Kishigami, P.~R. Krehbiel, V.~Kuzmin, Y.~J. Kwon, J.~Lan,
  R.~LeVon, J.~P. Lundquist, K.~Machida, K.~Martens, T.~Matuyama, J.~N.
  Matthews, M.~Minamino, K.~Mukai, I.~Myers, S.~Nagataki, R.~Nakamura,
  T.~Nakamura, T.~Nonaka, S.~Ogio, M.~Ohnishi, H.~Ohoka, K.~Oki, T.~Okuda,
  M.~Ono, R.~Onogi, A.~Oshima, S.~Ozawa, I.~H. Park, M.~S. Pshirkov,
  J.~Remington, W.~Rison, D.~Rodeheffer, D.~C. Rodriguez, G.~Rubtsov, D.~Ryu,
  H.~Sagawa, K.~Saito, N.~Sakaki, N.~Sakurai, T.~Seki, K.~Sekino, P.D. Shah,
  F.~Shibata, T.~Shibata, H.~Shimodaira, B.~K. Shin, H.~S. Shin, J.~D. Smith,
  P.~Sokolsky, R.~W. Springer, B.~T. Stokes, T.~A. Stroman, H.~Takai,
  M.~Takeda, R.~Takeishi, A.~Taketa, M.~Takita, Y.~Tameda, H.~Tanaka,
  K.~Tanaka, M.~Tanaka, R.~J. Thomas, S.~B. Thomas, G.~B. Thomson, P.~Tinyakov,
  I.~Tkachev, H.~Tokuno, T.~Tomida, S.~Troitsky, Y.~Tsunesada, Y.~Uchihori,
  S.~Udo, F.~Urban, G.~Vasiloff, T.~Wong, M.~Yamamoto, R.~Yamane, H.~Yamaoka,
  K.~Yamazaki, J.~Yang, K.~Yashiro, Y.~Yoneda, S.~Yoshida, H.~Yoshii, and
  Z.~Zundel.
\newblock Gamma ray showers observed at ground level in coincidence with
  downward lightning leaders.
\newblock {\em Journal of Geophysical Research: Atmospheres},
  123(13):6864--6879, 2018.

\bibitem{LaborStatistics}
Stanislava Ilic-Godfrey.
\newblock Artificial intelligence: taking on a bigger role in our future
  security.
\newblock {\em Beyond the Numbers: Employment \& Unemployment}, 10(9), May
  2021.

\end{thebibliography}
\bibliographystyle{unsrt}

\end{document}